# Cross-referencing social media and public surveillance camera data for disaster response

Chittayong Surakitbanharn, +Calvin Yau, +Guizhen Wang, +Aniesh Chawla, +Yinuo Pan,
+Zhaoya Sun, +Sam Yellin, +David Ebert, +Yung-Hsiang Lu, ++George K. Thiruvathukal
Stanford University, Stanford, CA 94305, USA. Phone: 858-923-9346
+Purdue University, West Lafayette, IN 47907, USA. Phone: 765-496-3747
++Loyola University, Chicago, IL 60626, USA. Phone: 773-508-8931
jao@stanford.edu, {yauc, wang1908, chawla9, pan125,
sun515, syellin, ebertd, yunglu}@purdue.edu, gkt@cs.luc.edu

*Abstract*—Physical media (like surveillance cameras) and social media (like Instagram and Twitter) may both be useful in attaining on-the-ground information during an emergency or disaster situation. However, the intersection and reliability of both surveillance cameras and social media during a natural disaster are not fully understood. To address this gap, we tested whether social media is of utility when physical surveillance cameras went off-line during Hurricane Irma in 2017. Specifically, we collected and compared geo-tagged Instagram and Twitter posts in the state of Florida during times and in areas where public surveillance cameras went off-line. We report social media content and frequency and content to determine the utility for emergency managers or first responders during a natural disaster.

## I. Introduction and Motivation

Social media (through smart-phones) and physical media (through the lowered costs of cameras) have changed disaster response and emergency management [1], [2]. Decision makers and first responders can now access crowd-sourced information via social media to obtain first-hand on-the-scene reports during a disaster [2]. In doing so, social media creates a channel for the public to convey both the urgency of the situation and specific needs for assistance.

Publicly available camera feeds can provide real-time and typically unaltered images from a hazardous area without risking lives [3]. However, surveillance cameras can be vulnerable to extreme weather. Their continued operation depends on a reliable electric infrastructure and network connection.

For example, Figure 1 shows a beach camera in Fort Lauderdale, FL at three different times during Hurricane Irma. The middle image shows the increasing severity of wind and rain in the 12 hours following the first image. However, the camera is off-line 12 hours later and has no value in obtaining on-the-ground visuals during and shortly after the storm hits.

Research has proposed the use of social media to augment emergency and disaster situation awareness [4], providing an alternative when surveillance cameras fail. Social media and network cameras can serve complementary roles: Social media reflects the observation, needs and sentiments of people; network cameras provide unbiased information. Social media can cover places where people visit; network cameras can continuously transmit data from remote locations and even after a city has been evacuated.

However, there has been a lack of rigorous demonstration to determine if social media posts accurately and reliably reflect a given situation. As a result, there remains uncertainty about the trustworthiness of social media and a lack of quantification of social media's utility during natural disasters and emergency events.

To bridge this knowledge gap, we propose and test a methodology to analyze the value (frequency and content) of social media in areas where surveillance cameras are off-line. To do so, we integrate real-time images and meta-data from network cameras with Instagram and Twitter posts during the 2017 Hurricane Irma in Florida. [4]. We aim to provide insight into the reliability and utility of social media data sources for emergency managers and first responders.

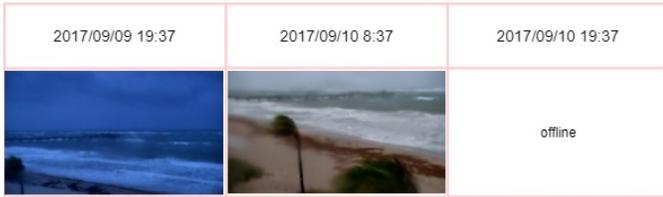

Fig. 1: A camera feed from Fort Lauderdale, FL during Hurricane Irma.

## II. RELATED WORK

Prior work has investigated both the use of network cameras and social media for emergency and disaster preparedness, citing that both can increase situation awareness. However, it is unclear whether social media can substitute the images and information provided by surveillance cameras when their feeds are disrupted.

### A. Images and Videos for Emergency Response and Public Safety

Images and videos are types of physical media that play crucial roles in emergency response, both in assessing the situations and conducting post-event (forensic) analyses [1]. News events are often accompanied by visual data from eyewitness' mobile phones (sometimes in real-time) [5]. Visual data from witnesses has advantages because the information reflects what people actually see. Nevertheless, such crowd sourced image and video data also has limitations. For example, it is difficult to perform "before-and-after" comparisons because it may be unlikely that people take photographs (or videos) at the same locations before, during, and after a news event occurs.

In contrast, network cameras can continuously acquire visual data at fixed locations, and the data can be used for comparison [6]. Taking an example from Houston traffic cameras during the 2017 Hurricane Harvey, Figure 2 (a) shows that Interstate 610 was flooded on August 28, 2017. Figure 2 (b) shows that at the same location the flood water has receded by August 30, 2017.

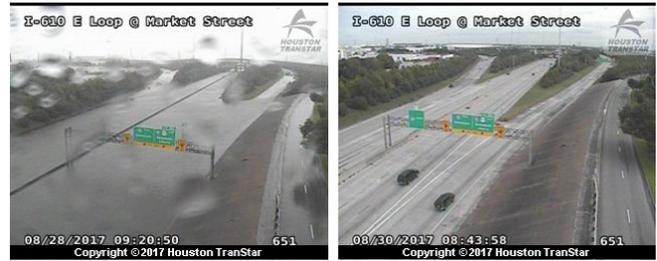

(a)        (b)

Fig. 2: The intersection of I-610 East and Market Street in Houston during Hurricane Harvey. (a) A flooded highway on August 28. (b) The flood has receded on August 30.

Prior work has explored using network cameras for public safety [7]. Recently, Alam and colleagues [3] demonstrated a prototype showing drivers road conditions from real-time images captured by traffic cameras. Su and colleagues [8] suggested creating a system that can harvest data from a wide range of network cameras, not limited to traffic cameras. However, the fusion of physical media (e.g., video feeds) and social media has yet to be explored, especially in the emergency and natural disaster domain. Thus, this paper extends previous work by integrating sources of social media streams with data and meta-data from public network cameras.

### B. Social Media and Natural Disasters

Social media has become integral to emergency response and disaster management in the past decade. The availability of social media facilitates multiple dimensions of communications, such as within or between citizen groups and disaster management authorities [2].

For example, social media platforms (e.g., Twitter, Facebook, Instagram) can be channels to provide updates to citizens about weather conditions or road closures. Conversely, social media can also allow citizens to provide disaster managers "on-the-ground" pictures and reports.

Previous research insinuates that increased information flow will improve disaster resilience and response [9]. However, it is still uncertain whether more data and information leads to better outcomes of public safety [10].

One challenge about social media is the verification of the time and accuracy of posted information. For example, a flood image taken by a user may not necessarily be posted to Twitter or Instagram at the time it was taken, nor may it be geo-tagged to the accurate location where it was taken. If first responders were to act on such information, resources during an emergency could be mis-allocated.

Furthermore, relevant information may be hidden among a slew of unrelated posts, which has led to studies about the content of social media during a crisis or disaster [11], [12]. Several studies have cited that social media can increase situation awareness [13], [14], but are focused on macro-level analyses (i.e., the overall quantity and sentiment of social media during an emergency event).

It is still uncertain whether social media can replace physical media focused on a particular geography (such as a network of cameras) to attain situation awareness during a natural disaster. To address this gap, we identify the locations and times network cameras went off-line during a hurricane and search social media to determine the frequency of disaster related content.

## III. Methodology

Our case study is Hurricane Irma making landfall in Florida, which began approximately on September 9, 2017. We take the following methodological steps to determine if social media is a viable alternative to network cameras during such events.

### A. Identifying Disruptions in Physical Media

First, we determine camera locations in Florida and identify a sample of cameras that went off-line during Hurricane Irma. Second, using the camera meta-data, we discover time intervals in which the selected cameras went off-line. Third, we define 2 mi × 2 mi, 10 mi × 10 mi, and 20 mi × 20 mi areas centered about the cameras that went off-line to search for social media posts.

### B. Physical Media to Social Media

In order to compare the social media information to that from surveillance cameras, we first collect social media posts during Hurricane Irma for Florida and the Caribbean. We use a visual analytic system (Figure 3) to aid in the collection and analysis of the social media data [15]. The social media collection system samples Twitter and Instagram posts, with a focus on geo-tagged data, or approximately 2% of total posts from these sources.

We then utilize the camera off-line disruption times (as temporal constraints) and our geographic boundaries to search the social media. In doing so, we report the total amount of Twitter and Instagram posts in our constrained areas. Afterwards, we report the total number of posts relevant to Hurricane Irma by searching for posts containing "Irma."

Finally, we analyze and report the total number in Twitter and Instagram that may provide on-the-ground information related to the event. For the content analysis, we visually inspect each post to determine if they provide live weather conditions or other information that would be of similar nature to surveillance cameras.

## IV. Results

This section reports the frequency and content of surrounding social media postings when network cameras were disrupted. Specifically, we focus on searching for images on Twitter and Instagram that can aid in emergency response.

### A. Visual Data from Network Cameras

We first use a camera network system to access 346 cameras in Florida and the Caribbean with live video feeds focused on public areas. After defining this network, we collect one image every ten minutes between September 6-12, 2017, resulting in over 210,000 total images. Although the resolutions vary from 270x150 to 1920x1080, most gave a general idea of weather.

Of these cameras, we find 35 cameras that went off-line (no data or repeating images) during Hurricane Irma. We use news sources to identify cities in Florida that were most affected by Hurricane Irma, and cross reference with the set of off-line cameras. We identify fifteen cameras in seven metropolitan areas that were disrupted (Table I). From this set, we remove two cameras due to a lack of social media data in their surrounding areas.

Based on the camera locations, we then generate geographic boundaries to search for social media posts. We apply the same geographic boundary for cameras with the exact same coordinates and disruption times.

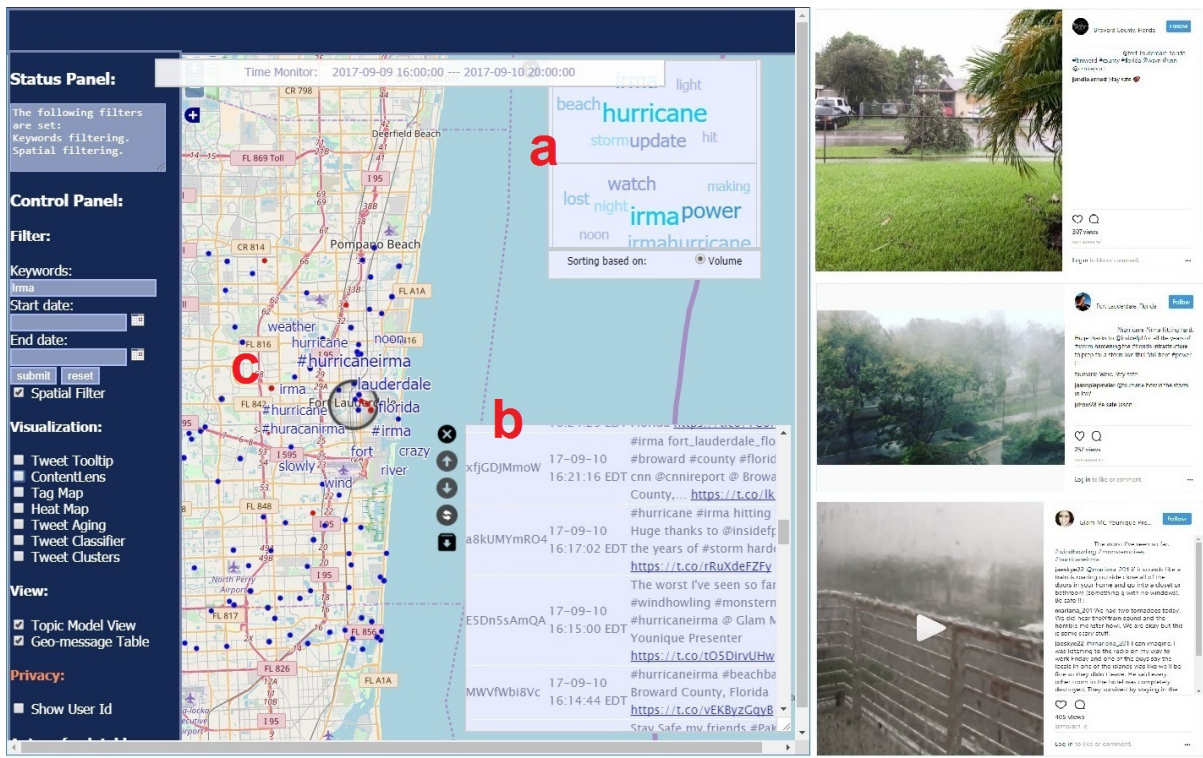

Fig. 3: An example system that displays social media around Fort Lauderdale during Irma. Residents posted images and videos of Irma outside their homes on Twitter (blue dots) and Instagram (red dots). Furthermore, (a) shows trending topics in the area, (b) shows content and meta data, and (c) shows prominent keywords and hashtags.

### B. Social Media Data

We utilize a dataset of of geo-tagged Twitter and Instagram posts that encompassed the area from 30°N,87°W to 24°N,79°W and spanned from 8/20/2017 to 9/13/2017. Within this total area and time period, a total of 8,800 geo-tagged Twitter and Instagram posts accessible by our system contained the term "Irma."

A total of 107, 1002, and 3089 Twitter and Instagram posts were found in the respective 2 mi × 2 mi, 10 mi × 10 mi, and 20 mi× 20 mi areas surrounding the thirteen cameras in seven unique locations. Of the total social media content, a respective of 37, 254, and 954 posts contained the term "Irma." Of the posts containing "Irma," a respective 36, 248, and 954 posts were images on Instagram. Of the Instagram images, a respective 11, 110, and 411 were found to contain information potentially relevant to disaster response. Additional details about the frequency of social media posts can be found in Table I.

### C. Content Analysis

The social media images we find relevant (i.e., showing weather) in our visual analysis were 10.3% in 2 mi × 2 mi, 11.0% in 10 mi × 10 mi, and 13.3% in 20 mi× 20 mi. At the 20 mi× 20 mi catchment scale, Miami contributed the most social media posts (79.8%) while Naples contributed the least (0.32%). Figure 4 provides an example of network camera images before going off-line (top row) and the types of Instagram images and video (bottom row) that we consider relevant in our visual analysis. All posts and images occur during the respective camera disruptions.

In the St. Augustine example (Figure 4(a)), the camera image of a flooded parking garage was posted to Instagram on 9/11/2017 at 0:26, approx-imately 0.97 miles away from the camera location. In the Naples example (Figure 4(b)), an image of a fallen tree was posted to Instagram on 9/11/2017 at 20:26, approximately 5.48 miles away from the camera location. In the Fort Myers example (Figure 4(c)), a video showing a tree falling and hitting a car was posted to Instagram on 9/10/17 at 15:42, approximately 4.44 miles away from the camera location.

## V. Discussion

We introduce a methodology and provide an analysis about how social media can be harnessed to gain better geographical and temporal situation context during a natural disaster. We use this section to make inferences from our findings, discuss analytical limitations, and guide future work.

### A. The Validity of Social Media

We find that 10% to 13% of Instagram images can provide relevant information in a 20 mi× 20 mi space, which can lead to several interpretations. Most social media posts are not relevant to attaining emergency or disaster information since they do not reflect conditions related to the event. This supports the need for a directed filtering and searching functions if an emergency manager wants to employ social media for situation awareness.

In our study, we only utilize the term "Irma" and does not include an array of other disaster related terms like "hurricane," "flood," or "disaster." We decide to use only one search term since our pilot studies demonstrate that posts that contained "Irma" is highly likely to contain other disaster related terminology, but the opposite does not hold true. For example, we want to exclude false positive posts such as "my meal is a disaster." Future work can further test the sensitivity posts to different search terms and arrays, which may provide emergency managers guidance for querying social media.

There are several factors that effects the quantity of collected social media posts. For example, the search areas for the cameras surrounding Fort Myers overlap and lead to the recounting of some posts.

Additionally, Hurricane Irma was the largest evacuation in Florida's history, which may have influenced certain areas, like Naples or St. Augstine, to have lower posting frequency. A future study that correlates social media posts to physical damage, socio-demographic factors, or digital connectedness could help define areas where social media may have more (or less) utility in emergency response.

We find social media posts in Naples and St. Augustine that could potentially provide first responders on-the-ground information within a one and five mile distance. For example, the social media images shown in Figure 4 could help first responders estimate environmental damages nearby.

Both surveillance cameras and social media rely on an operating power infrastructure. Like any system, network cameras are subject to many reasons of failures, such as loss of power, loss of network connection, or being blown away by strong winds. There is no systematic study of the reliability of network cameras, which is a crucial step before relying on network cameras in emergencies. Furthermore, incorporating power outage and other infrastructure data could also help understand the effect other systems have on social media.

It is unclear why most posts we find searching the term "Irma" are on Instagram rather than Twitter. We anticipate that access to the "firehose" of social media data from both Twitter and Instagram may affect the frequency of posts and potentially the Twitter/Instagram ratio. However, a larger dataset may also introduce a data overload for an end-user, which warrant future studies about how both physical and social media information is perceived by a domain user. Furthermore, a larger dataset may require image recognition or more advance filtering technologies to identify valuable information rather than manual visual inspection.

We did not identify whether the location and content of the posts are accurate. For example, we do not know if the image of the flooded garage or tree falling actually corresponds to the geo-tagged locations. Nor do we know whether the person posting is witnessing the event first-hand or re-posting an image they found elsewhere online. The question of reliability requires further investigation to determine the utility of cross-referencing physical and social media.

## VI. Conclusion

Social media has been widely used in emergencies. This article provides evidence that social media may be an alternative information source when network cameras fail to provide real time information.

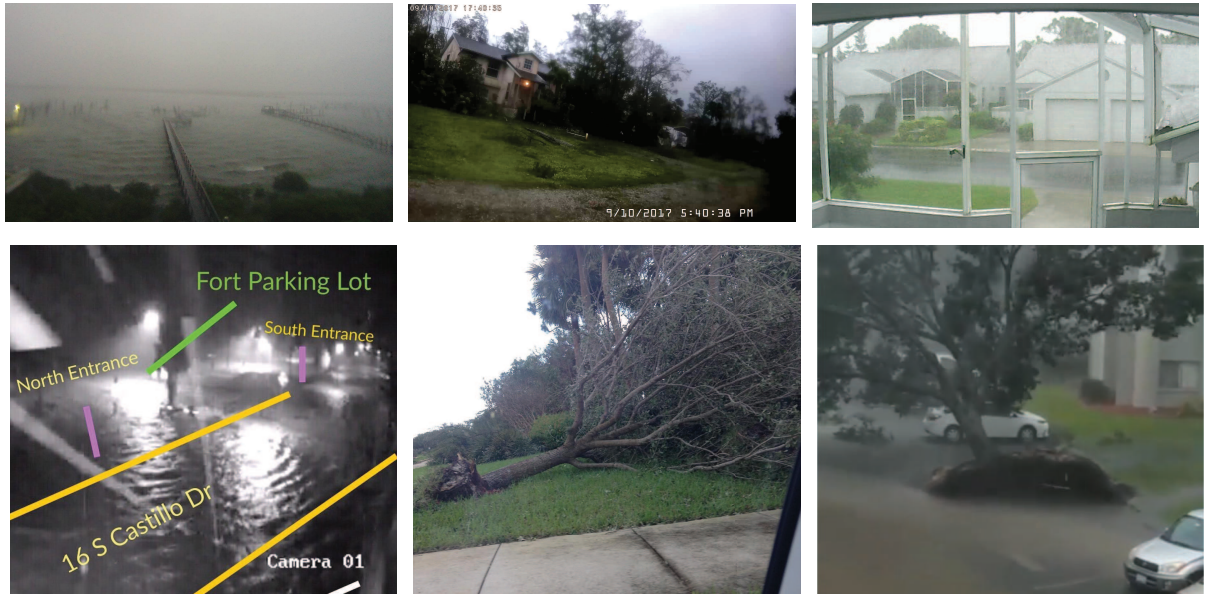

Fig. 4: A comparison of physical network camera images (top row) prior to disruptions, Instagram posts (bottom row) during network camera disruptions, post content, and metropolitan area.

This paper introduces and tests a methodology to determine the frequency and content of Twitter and Instagram posts during times and in areas that network cameras failed using Hurricane Irma as a case study. Even though many studies have considered using social media data, the integration and cross referencing of data with physical media is often overlooked. From this research, we gained a better understanding of relationships between disrupted network cameras and social media during a natural disaster. Through filtering and manual image content analysis, we determined the fidelity of which social media content substitute surveillance cameras in similar areas.

Machine learning techniques may scale our methodology to larger datasets. Manual filtering (as done here) and subsequent labeling are the first steps in future machine learning based solution. In future work, we may test the effectiveness of natural language processing to filter posts related to disasters as well as investigate the reliability of social media posts.

Finally, future work will require a synthesis of both network cameras and social media to observe and respond to emergencies. As cameras become widely used in the Internet of Things and social media becomes more prevalent, further investigation about content, data noise, and disruptions are needed to understand the utility of vast amounts of real-time data in emergencies.

APPENDIX A

| City | Camera Latitude | Camera Longitude | Disruption Start time | Disruption End time | Cameras in Group | # Posts | | | # Posts Filtered by "Irma" | | | # Filtered Instagram Posts | | | # Relevant Instagram Posts | | |
|---|---|---|---|---|---|---|---|---|---|---|---|---|---|---|---|---|---|
| | | | | | | 2mi × 2mi | 10mi × 10mi | 20mi × 20mi | 2mi × 2mi | 10mi × 10mi | 20mi × 20mi | 2mi × 2mi | 10mi × 10mi | 20mi × 20mi | 2mi × 2mi | 10mi × 10mi | 20mi × 20mi |
| Naples | 26.26 | -81.62 | 9/10/17 17:45 | 9/12/17 2:45 | 2 | 0 | 0 | 10 | 0 | 0 | 7 | 0 | 0 | 7 | 0 | 0 | 5 |
| Key West | 24.55 | -81.78 | 9/9/17 20:55 | 9/12/17 2:45 | 1 | 81 | 90 | 92 | 24 | 24 | 24 | 24 | 24 | 24 | 6 | 6 | 6 |
| St. Augustine | 29.89 | -81.29 | 9/10/17 23:25 | 9/12/17 19:56 | 1 | 3 | 51 | 82 | 1 | 5 | 15 | 1 | 3 | 13 | 1 | 3 | 9 |
| Fort Myers | 26.67 | -81.79 | 9/10/17 11:44 | 9/12/17 1:44 | 2 | 0 | 55 | 216 | 0 | 29 | 55 | 0 | 28 | 52 | 0 | 15 | 26 |
| Fort Myers | 26.69 | -81.87 | 9/10/17 14:04 | 9/12/17 1:44 | 1 | 0 | 53 | 137 | 0 | 30 | 59 | 0 | 30 | 58 | 0 | 17 | 31 |
| Tampa | 28.06 | -82.41 | 9/12/17 5:16 | 9/12/17 19:56 | 1 | 3 | 33 | 288 | 1 | 4 | 43 | 1 | 4 | 40 | 0 | 1 | 6 |
| Miami | 25.78 | -80.34 | 9/9/17 14:15 | 9/11/17 3:05 | 5 | 20 | 720 | 2264 | 11 | 162 | 751 | 10 | 159 | 751 | 4 | 68 | 328 |
| Total | | | | | 13 | 107 | 1002 | 3089 | 37 | 254 | 954 | 36 | 248 | 945 | 11 | 110 | 411 |

TABLE I: Camera disruption times and social media counts for Hurricane Irma around seven camera locations.

BIOGRAPHY OF THE PRESENTING AUTHOR

**Aniesh Chawla**

M.S., Electrical and Computer Engineering,

Purdue University, West Lafayette IN.

The Continuous Analysis of Many CAMeras ($CAM^2$)

Website: *https://www.cam2project.net/*

Email: chawla9@purdue.edu